\documentclass[a4paper,preprint]{aastex}
\usepackage{graphicx}
\usepackage{natbib}

\newcommand{\bh}{{\rm bh}}
\newcommand{\apgt}{\ {\raise-.5ex\hbox{$\buildrel>\over\sim$}}\ }

\begin{document}

\title{Tidal Disruption of a Solar Type Star by a Super-Massive Black Hole} 

\author{
  Shai Ayal\altaffilmark{1,2}, 
  Mario Livio\altaffilmark{2}, and
  Tsvi Piran\altaffilmark{1}}

\altaffiltext{1}{Racah Institute of Physics, Hebrew University,
  Jerusalem 91904, Israel} 

\altaffiltext{2}{Space Telescope Science Institute, 3700 San Martin
  Drive, Baltimore, MD 21218, USA}

\begin{abstract}
  We study the long term evolution of a solar type star that is being
  disrupted by a super massive ($10^6$~M$_\odot$) black hole. The
  evolution is followed from the disruption event, which turns the
  star into a long thin stream of gas, to the point where some of this
  gas returns to pericenter and begins its second orbit around the
  black hole. Following the evolution for this long allows us to
  determine the amount of mass that is accreted by the black hole. We
  find that approximately 75\% of the returning mass is not accreted
  but instead becomes unbound, following the large compression
  characterizing the return to pericenter.  The impact of a tidal
  disruption on the surrounding gas may therefore be like that of two
  consecutive supernove-type events.
\end{abstract}

\section{Introduction}
\label{sec:Intro}

It has long been suggested that supermassive black holes (with a black
hole mass $M_\bh\sim10^6$~M$_{\odot}$) in relatively low-luminosity
active galactic nuclei (AGNs) can be fed by tidal disruptions of stars
that are on nearly radial orbits
\citep[e.g.][]{1978MNRAS.184...87F,1982ApJ...262..120L,1985MNRAS.212...23C,1988Natur.333..523R,1990Sci...247..817R}.
The frequency of such events is expected to be of the order of
$10^{-4}$~yr$^{-1}$ in a galaxy like M31.  In AGNs with more massive
central black holes ($M_\bh\ga10^8$~M$_{\odot}$), stars are
``swallowed'' whole, since the ratio of the event horizon radius to
the tidal disruption radius increases with the black hole mass
$\left(\sim M_\bh^{2/3}\right)$.

Although direct observational evidence for the disruption process is
lacking, a few observations have been tentatively identified as being
associated with disruption events. For example, an ultraviolet flare
at the center of the elliptical galaxy NGC~4552, although rather weak,
has been suggested to result from such an event \citep{Renzini95}.
Other sudden eruptions that have been interpreted as potentially
resulting from tidal disruptions include outbursts in IC 3599, NGC
5905, RXJ 1242.6-1119, RXJ 1624.9+7554, and RXJ 1331.9-3243
\citep[e.g.][and references therein]{Komossa99b,Komossa99a,Grupe99}.
It has also been suggested that the sudden appearance and variability
of the double-peaked Balmer lines in the active nucleus NGC~1097 is at
least broadly consistent with being produced in a ring resulting from
tidal disruption \citep{1995ApJ...443..617S}. Similarly, it has been
proposed that an outburst observed in the Seyfert galaxy NGC~5548 was
due to a single star falling into a $\sim10^7$~M$_{\odot}$ black hole
(\citealt{Peterson86}; although other interpretations exist, e.g.\ 
\citealt{Terlevich88,Kallman88}). More recent work on the broad
emission line region of this galaxy, although interesting in its own
right, did not shed any new light on the cause of the variability
\citep{Goad98}.

Several aspects of the problem of stellar encounters with supermassive
black holes and of tidal disruption have been examined both
analytically and numerically
\citep[e.g.][]{1982ApJ...263..377N,1983ApJ...273..749B,1983A&A...121...97C,Hills88,1992ApJ...385..604K,1994ApJ...422..508K,1992MNRAS.255...92S,1993MNRAS.260..463S,1993ApJ...410L..83L,Marck96}.
In addition, the interaction of a white dwarf with a massive black
hole has been studied both generally
\citep[e.g.][]{1994ApJ...432..680F} and in the context of gamma-ray
bursts \cite[e.g.][]{1999ApJ...520..650F}.

In the present work, we follow the evolution of a star that is tidally
disrupted. In particular, we calculate the properties of the
disruption debris to longer times than in some of the previous
studies, using a Post-Newtonian, Smooth-Particle-Hydrodynamics (SPH)
code.

The numerical method is described in $\S2$, the results are presented
in $\S3$, and a discussion and conclusions follow.

\section{The Numerical Method}
\label{sec:NumMeth}

We first introduce some notation that will be used throughout the
paper. The mass of the black hole is denoted by $M_\bh$. The
gravitational radius of the black hole is $R_g=2GM_\bh/c^2$. To
measure the strength of the tidal encounter we use the dimensionless
parameter \citep[e.g.][]{1977ApJ...213..183P}
\begin{equation}
  \label{eq:1}
  \eta_t = 
  \left(\frac{M_*}{M_\bh}\frac{R_p^3}{R_*^3}\right)^{1/2}~~,
\end{equation}
and to measure the magnitude of the relativistic effects we use 
\begin{equation}
  \label{eq:3}
  \eta_r =
  \frac{R_p}{R_g}~~,
\end{equation}
where $M_*$ and $R_*$ are the mass and radius of the star respectively 
and $R_p$ is the radius at the pericenter.

Simulating the full evolution of a star that is tidally disrupted by a
black hole is a challenging numerical task. Once passing the
pericenter, the star is tidally disrupted into a very long and dilute
gas stream.  As noted already by \cite{1994ApJ...422..508K}, each
fluid element in this stream follows an almost test particle orbit.
The fraction of the gas that is bound to the black hole eventually
returns to pericenter and moves on to start another orbit.
Relativistic orbital precession can cause this outgoing gas stream to
collide with the incoming stream. Clearly, this problem consists of
highly varying spatial configurations of matter, with much ``empty
space.'' These attributes led to our selection of SPH as the numerical
scheme used to tackle this problem. Indeed, previous works that used
grid based codes (e.g.\ 
\cite{1993ApJ...418..163K,1993ApJ...418..181K,1994ApJ...432..680F,1997ApJ...479..164D})
only followed the star close to the pericenter. This is also true of
previous works by \cite{1989ApJ...346L..13E}, and
\cite{1993ApJ...410L..83L} that used SPH.

We use the (0+1+2.5) Post-Newtonian (PN) SPH code described in
\cite{sphpn}.  This code implements the formalism of Blanchet, Damour
\& Sch\"afer (1990; hereafter BDS) and features full Newtonian gravity
and hydrodynamics (0PN), the first order effects of general relativity
on the gravity and hydrodynamics [known as the first post-Newtonian,
(1PN) approximation], and gravitational wave damping (2.5PN).  The
independent matter variables used in this formalism consist of the
following set: $\rho_*$ the coordinate rest mass density,
$\varepsilon_*$ the coordinate specific internal energy and ${\bf w}$
the specific linear momentum. In fully relativistic terms these are
defined as:
\begin{eqnarray}
  \rho_*  &=& \sqrt{g}u^0\rho,\\
  \varepsilon_* &=& \varepsilon(\rho_*),\\
  w_i &=& \left(c^2 + \varepsilon + p/\rho\right)\frac{u_i}{c},
\end{eqnarray}
where $\rho$ is the rest mass density, $\varepsilon(\rho)$ is the
specific energy, $p(\varepsilon,\rho)$ is the pressure and $u^\mu$ is
the four-velocity (Greek indices run from 0 to 4, latin indices from 1
to 3). The corresponding BDS variables are the above quantities
neglecting all terms except 0PN, 1PN and 2.5PN. Using these variables
the formalism yields an evolution system which consists of 9~Poisson
equations and 4~hyperbolic equations which we solve as explained in
\cite{sphpn}.

We model the black hole using a massive point particle that has no
hydrodynamical interactions.  In order to be consistent with the 1PN
approximation we must ensure that all the relativistic effects are
small (of the order of 10\%).  This excludes simulating the strong
field regions near the black hole's horizon. We enforce this limit
using a fully absorbing boundary condition at $10R_g$. Every particle
crossing into this region is assumed to be accreted and taken out of
the simulation. Since the mass of the star is negligible compared the
mass of the black hole, we do not need to increase the mass of the
black hole for every particle crossing the $10R_g$ boundary.

The statistical nature of SPH does not handle well single, separate
particles. Indeed, the entire formalism is built on the assumption
that each particle interacts hydrodynamically with about 60 (in 3D)
other particles at all times. This number of interactions is needed to
form a good sample of the fluid properties for an accurate calculation
of gradients. In order to maintain this number of interactions under
varying conditions, most SPH codes employ adaptive smoothing lengths
\citep[e.g.][]{benz_rev,mon_rev}, in effect changing the resolution at
each point. In our problem, we have widely varying length scales when
the gas stream expands, where varying the smoothing length is
advantageous.  On the other hand, the difference in the particle
eccentricities causes them to separate when approaching pericenter,
and they arrive there almost one by one. At this stage, maintaining
hydrodynamic interaction with 60 other particles would require huge
smoothing lengths, where a small change in the smoothing length would
lead to a large change in the number of interactions. This, coupled
with an algorithm that tries to maintain a fixed number of
interactions, can cause large oscillations in the smoothing length,
which in turn introduce a highly varying number of hydrodynamic
interactions. These huge smoothing lengths for particles approaching
pericenter also mean that we have a very coarse resolution at a
crucial stage.

In order to overcome this numerical difficulty we introduce a
``particle splitting'' (PS) scheme. Whenever a particle satisfies some
splitting criterion we split it into new particles each having a
smaller mass and smoothing length so that the overall mass is
conserved. Thus the PS algorithm consists of two parts---the splitting
criterion and the splitting method. In the method we use, maximal
splitting, we split each particle into 13 particles, giving the
original particle 12 new neighbors (12 is the maximum number of
spheres that can ``touch'' a sphere with the same radius). Another
possible method is minimal splitting (while still maintaining a quasi
spherical symmetry)---we split each particle into~5, adding 4
particles at the edges of a tetrahedron centered on the original
particle. We found that the minimal splitting method tends to produce
lumpier particle distributions, we therefore used maximal splitting in
both the PS runs presented in this work.  We split each particle into
13 particles, each having half the smoothing length of the original
particle, spaced by one original smoothing length. This splitting
method conserves the particle's interaction radius which is twice it's
smoothing length. As a splitting criterion we used the ratio between
each particle's smoothing length and the average smoothing length.
Whenever the smoothing length of a bound particle exceeds twice the
average smoothing length we split this particle. We split only bound
particles so as not to waste computational time on the unbound debris,
in which we are not interested in the present work. In order for the
unbound debris smoothing length not to dominate the average smoothing
length which we use for the splitting criterion, we enforce a maximum
smoothing length of twice the average smoothing length for unbound
particles.  This causes unbound particles of gas to have fewer and
fewer hydrodynamical interactions, and their motion to be dominated by
gravity, as is expected. Another computational time reducing technique
that we use is to delete any particle that is both unbound and is
farther away than $2500R_g$. The latter criterion ensures that the
deletion of these particles does not affect the dynamics near
pericenter.

In order to estimate the errors in using the particle splitting method,
we compared the results of three runs. The first run was a conventional 
SPH run with a fixed number of 4295 particles, denoted by F1. The 
second and third runs were PS runs denoted by PS1 and PS2 and they 
differ only in the number of particles at the initial time. These parameters 
are summarized in Table~\ref{tab:parms}.
\def\en{\enspace}
\begin{table}[h]
\caption{The parameters of the three runs. F1 has a fixed number
      of particles, PS1 and PS2 are particles splitting runs differing
      only in the number of particles.}
  \begin{center}
    \begin{tabular}{|c|c|c|c|}\tableline
      Run & Type & Initial \# of particles & Maximum \# of particles\\ \tableline
      F1  & fixed & 4295 & 4296 \\
      PS1 & PS    & \en579  & 1528 \\
      PS2 & PS    & 1079 & 3607\\\tableline
    \end{tabular}
        \label{tab:parms}
  \end{center}
\end{table}
 
\section{Results}
\label{sec:Res}
\begin{figure}
  \begin{center}
    \includegraphics[width=10cm]{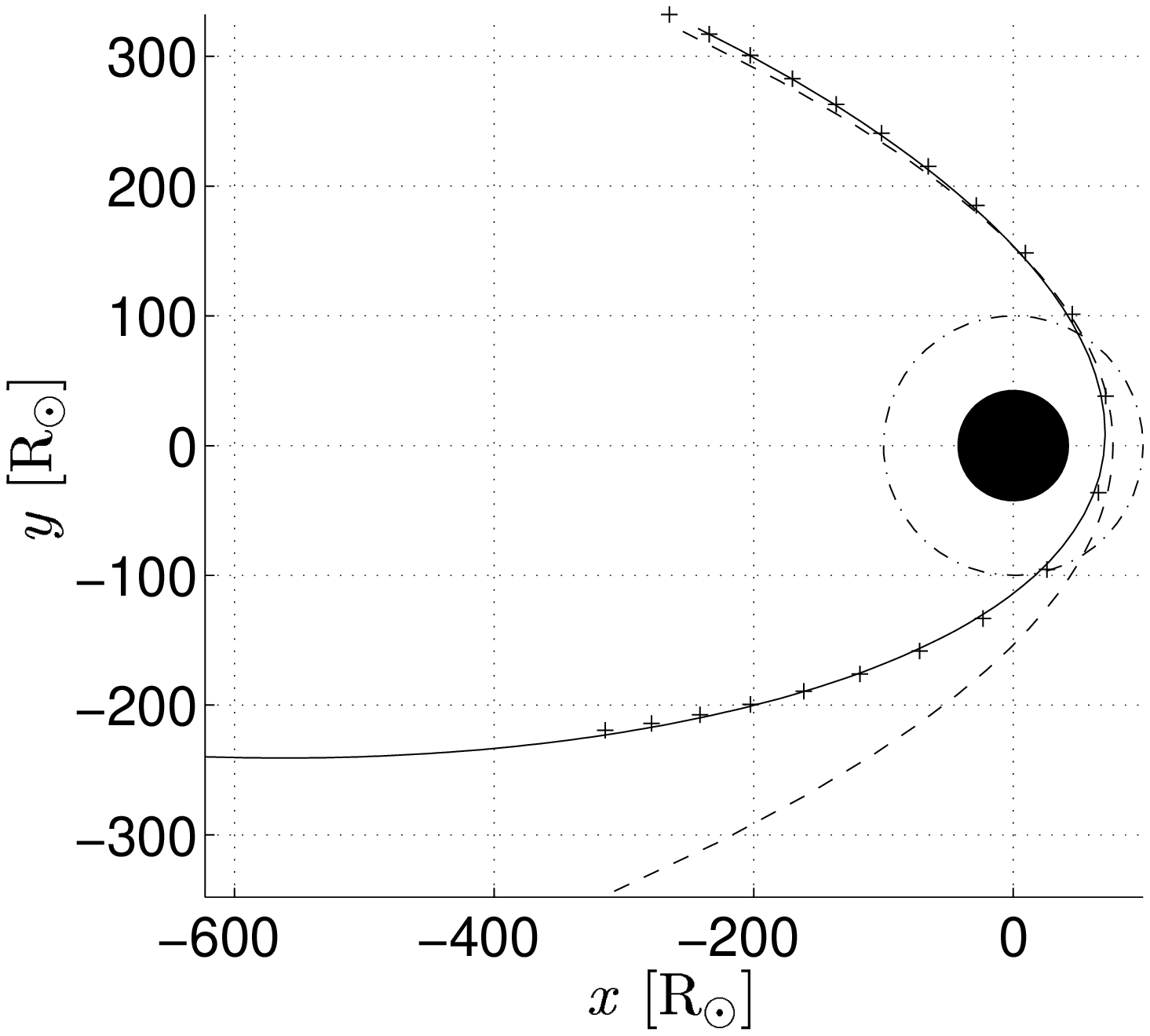}
    \caption{\label{fig:cm}
      The initial encounter: The star's CM position near pericenter
      (crosses), compared with the relativistic orbit (solid line) and
      Newtonian orbit (dashed line) with the same initial conditions.
      The black circle in the center has a radius of $10R_g$. The
      dash-dotted line is at the tidal radius. The star's CM is closer
      to the relativistic orbit, but it is not precisely on it.  }
  \end{center}
\end{figure}

As initial conditions we took a $\Gamma=5/3$ polytrope with a solar
radius and mass. This star was then put at a distance of $100R_g$ and
given the velocity of an appropriate parabolic Keplerian orbit with a
pericenter at $20R_g$. Giving the initial conditions at such a
distance ensures that any relativistic effects (at that position) are
negligible so that using Newtonian expressions for various quantities
(such as energy) is justified.  As can be seen in Fig.~\ref{fig:cm},
the star's center of mass (CM) follows an almost relativistic orbit,
which validates the use of the 1PN approximations for this problem.
For this run we obtained the values of $\eta_r\approx 18$ and
$\eta_t\approx 0.65$ at the pericenter. These values ensure that on
one hand the 1PN approximation is valid, and on other, that the tidal
interaction is sufficiently strong to lead to disruption.

We can roughly divide the disruption process into 2 qualitatively
different regimes. In the first, the tidal forces dominate and
the star is destroyed. This is followed, at about $t=8$ hours, by the
post disruption regime. This latter stage consists of the gas stream
phase and the accretions phase. During the gas stream phase the
pressure is negligible and the particles move on almost Keplerian
orbits. The accretion phase starts at $t=15$ days when the first
particles return to pericenter.

\subsection{Disruption}
The disruption phase for a solar type star was studied, using
Newtonian physics, by \cite{1989ApJ...346L..13E}. Later it was studied
in greater detail by \cite{1993ApJ...418..163K,1993ApJ...418..181K};
and by \cite{1997ApJ...479..164D} who included a general relativistic
treatment of the tidal potential of a Kerr black hole and Newtonian
hydrodynamics for the star. This phase was also studied by
\cite{1993ApJ...410L..83L} using Newtonian hydrodynamics on a fixed
Kerr background. We compare our results for this phase with these
previous studies.

\begin{figure}
  \begin{center}
    \includegraphics{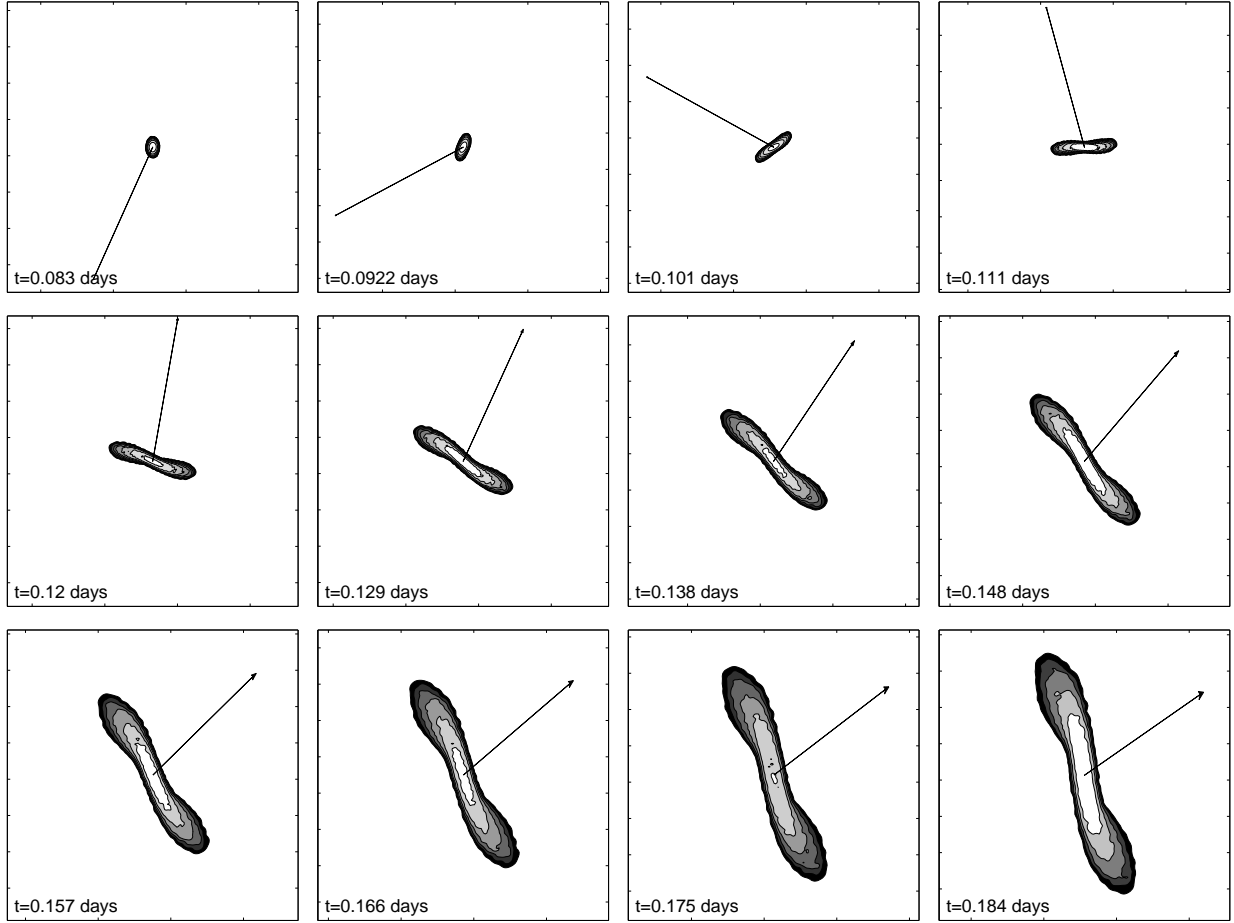}
    \caption{\label{fig:distcont}
      Contours of the rest-mass density on the orbital plane during
      the disruption. The arrow points to the BH. Time advances
      from left to right, top to bottom. All figures are to the same
      scale of 40~R$_\odot$ per side.}
      \end{center}
\end{figure}

\begin{figure}
  \begin{center}
    \includegraphics[width=8cm]{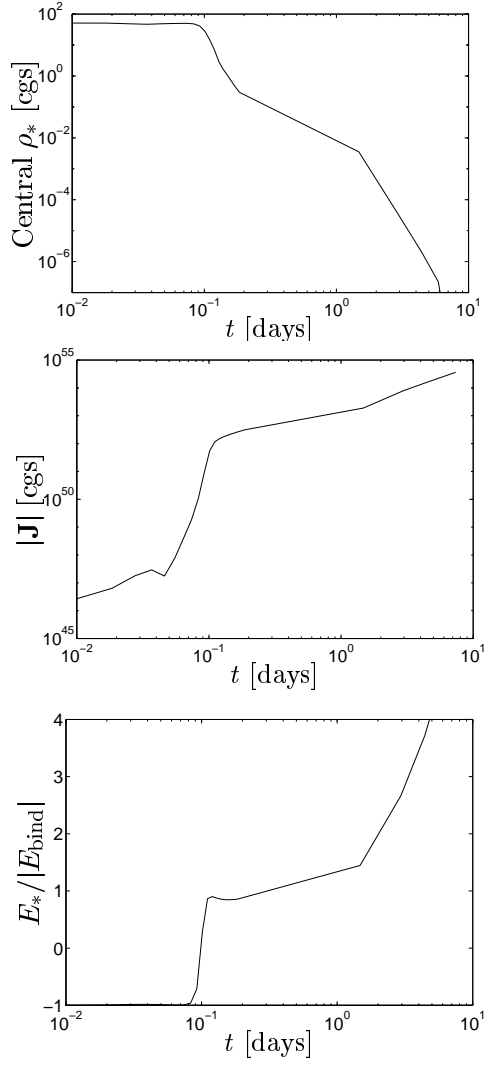}
    \caption{\label{fig:rsL}
      Central coordinate rest mass density $\rho_*$, angular momentum
      $|{\bf J}|$ (relative to the star's CM), and total energy
      of the star as a function of time.}
\end{center}
\end{figure}

The results shown are taken from run F1 which has a higher resolution
at this stage. The rest-mass density contours presented in
Fig.~\ref{fig:distcont} show the effect of the tidal forces on the
star.  In Fig.~\ref{fig:rsL} we show the central coordinate rest-mass
density $\rho_*$, the angular momentum $|{\bf J}|$, and total energy
$E_*$ of the star, during the disruption. The value of $\rho_*$ does
not increase beyond the initial value, and it falls rapidly after the
disruption.  Comparing to \cite{1993ApJ...410L..83L}, we find that our
encounter is somewhere between a non-relativistic encounter and a
single-compression relativistic encounter, as expected. The values of
$\rho_*$ are also close to those in \cite{1997ApJ...479..164D} and
\cite{1993ApJ...418..181K}. The total angular momentum relative to the
CM of the star increases by about 5~orders of magnitude at the
disruption, and a more gentle increase occurs afterwards, caused by
the gravitational torques acting on the debris.  The total energy of
the star also has a sharp increase at the disruption as the star
becomes unbound, followed by another rise as the star moves away from
the BH.

\begin{figure}
  \begin{center}
    \includegraphics[width=8cm]{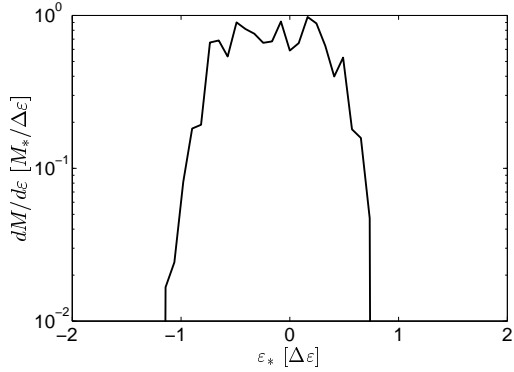}
    \caption{\label{fig:dmde}
      Differential mass distribution in specific energy for the
      disruption debris}
      \end{center}
\end{figure}

The differential mass distribution in specific energies
$dM/d\varepsilon_*$ is shown in Fig.~\ref{fig:dmde}. We use
$\Delta\epsilon=GM_\bh R_*/R_p^2$, the change in the BH potential
across the star \citep{1982ApJ...262..120L} as our energy scale. The
mass distribution is almost constant as predicted by
\cite{1988Natur.333..523R}. A comparison with
\cite{1989ApJ...346L..13E}, who calculated this quantity for the
Newtonian case, shows that in our calculation the width of the distribution is
smaller by approximately $0.5\Delta\epsilon$ (FWHM).

In general, we find that the 1PN energy $E_{\rm pn}$ is conserved to
better than 2\%. The energy radiated by gravitational radiation
during the disruption is negligible (compared to the total), amounting to 
$1.6\times10^{46}$~erg.

\subsection{Post Disruption}

\begin{figure}
  \begin{center}
    \includegraphics[height=10cm]{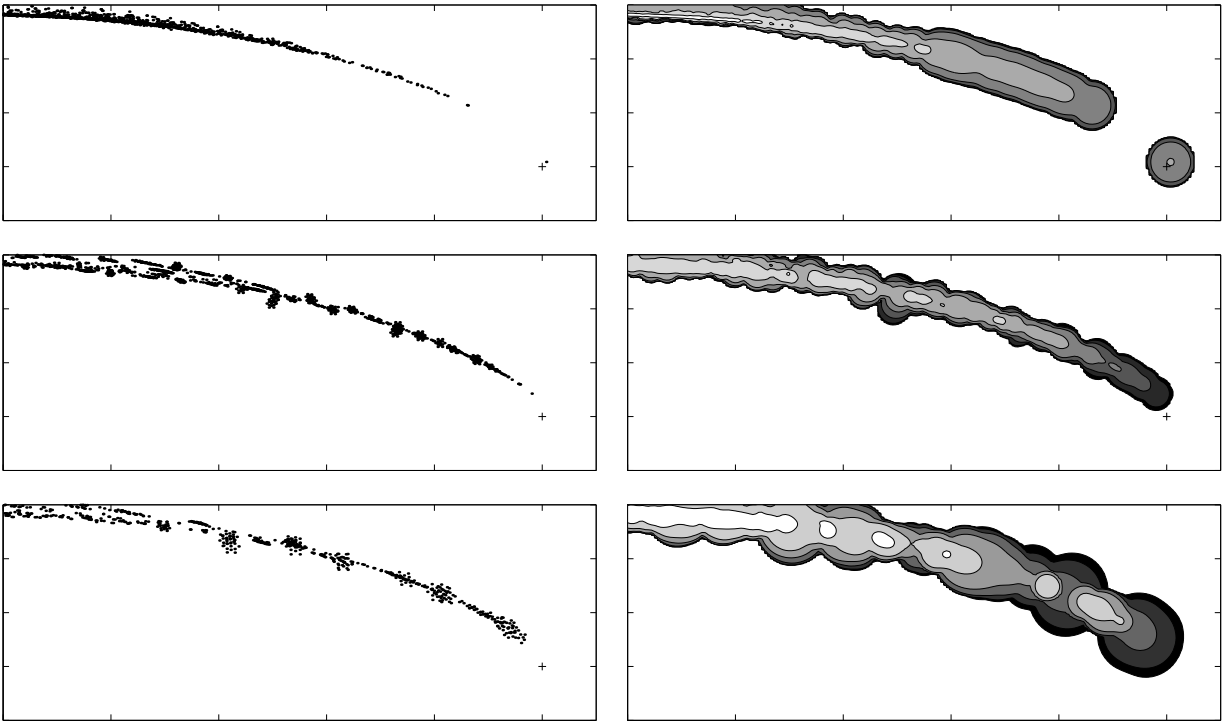}
    \caption{\label{fig:stpl}
      Stream at the beginning of second pericenter passage. On the
      left, the particle distribution, on the right the density
      contours. We show the three runs: on the top is the F1 run, in
      the middle the PS2 run and on the bottom the PS1 run. The width
      of the plotted area is $10^4$~R$_\odot$. Note that in both the
      PS runs, the {\em particle} density near pericenter is larger
      than in the F1 run. Also note the isolated particles approaching 
      pericenter in the F1 run, a phenomenon which is absent in the
      PS runs. The clumping in the particles is produced by the PS
      scheme.  }
      \end{center}
\end{figure}

The post disruption phase begins at about $t=8$~hours. At this stage
the gas is sufficiently dilute and cold to make the pressure very
small.  Consequently, the gas elements in the stream follow almost
exact geodesics.  A previous study on the gas stream phase, by
\cite{1994ApJ...422..508K}, used the thinness of the stream to
decouple the transverse properties of the stream from the variations
along its length. This same thinness introduces difficulties into the
numerical approach. The spherical nature of the SPH particles causes
the code to overestimate the stream width. This effect can be overcome
by using more particles but the increase in computational time renders
this approach impractical with current hardware.  As the SPH particles
approach pericenter for the second time, the PS method therefore
becomes essential. Without PS, the SPH particles approach pericenter
almost one by one and the strong compressional effects are manifested
through two- and three-particle interactions near the pericenter. This
small number of interactions could reduce the reliability of the
results. By using PS however, we increase the number of particles
approaching pericenter, thus increasing the number of interactions and
thereby ensuring that the hydrodynamical interactions remain adequate.
The effects of PS can be seen in Fig.~\ref{fig:stpl} where the
conventional run has a very non-uniform particle density even when
compared to the PS1 run (with only 1350 particles at this stage).

\begin{figure}
  \begin{center}
    \includegraphics[width=12cm]{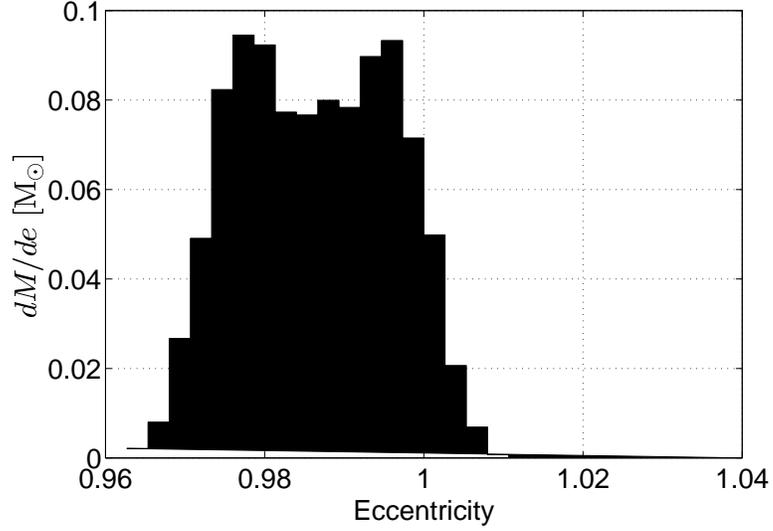}
    \caption{\label{fig:st_ec}
      The eccentricity distribution of the gas in the stream just
      before the first particle approaches pericenter (at $t\approx
      15\rm days$). The mean eccentricity is $\approx 0.994$. }
  \end{center}
\end{figure}

\begin{figure}
  \begin{center}
    \includegraphics[width=10cm]{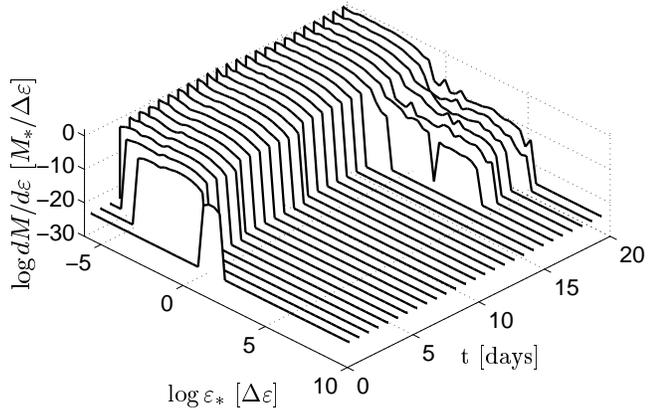}
    \caption{\label{fig:dmdesurf}
      Differential mass distribution in specific {\em internal} energy
      for the disruption debris as a function of time.}
  \end{center}
\end{figure}

\begin{figure}
  \begin{center}
    \includegraphics[width=8cm]{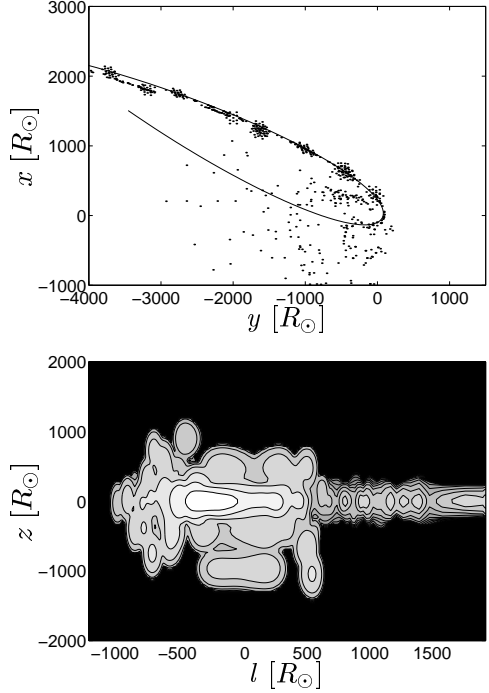}
    \caption{\label{fig:press}
      Pressure along a path. At the top we show the path, chosen to
      pass through the gas stream and pericenter. At the bottom we
      show logarithmic contours of the pressure. The contours are
      spaced by one decade. $l$ is the length parameter of the path,
      $l=0$ corresponds to the point where the path intersects the $x$
      axis, near pericenter and $l$ grows counter-clockwise.  }
      \end{center}
\end{figure}

\begin{figure}
  \begin{center}
    \includegraphics[width=10cm]{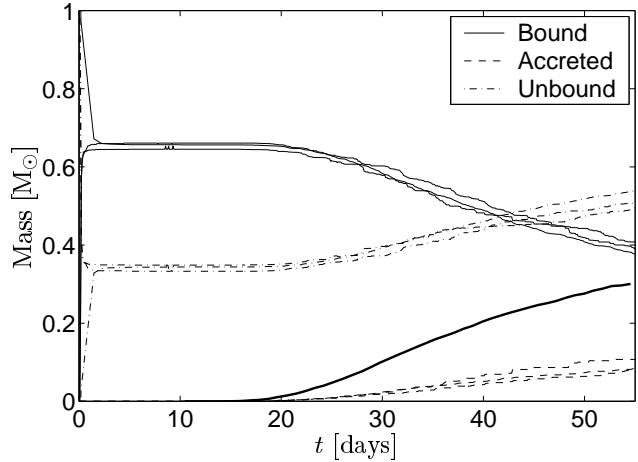}
    \caption{\label{fig:mb}
      Bound, accreted and unbound mass as a function of time. For each
      type of mass there are two lines for the two runs. As can be
      seen, the maximal difference is $\approx 10$\%. The heavy solid
      line is the accreted mass as predicted by assuming that all the 
      returning mass is accreted.}
  \end{center}
\end{figure}

In Fig.~\ref{fig:st_ec} we show the distribution of eccentricities for
the SPH particles in the gas stream phase. The mean eccentricity is
about 0.994, as the star was initially marginally bound. The
distribution does not change by much until the end of the stream
phase, when the first particle approaches pericenter.  The
differential mass distribution in specific internal energy as a
function of time is shown in Fig.~\ref{fig:dmdesurf}.  The earliest
time shown is close to that of Fig.~\ref{fig:dmde}. As can be seen, at
the gas stream phase the internal energy distribution shifts towards
lower energies, while the second passage through pericenter heats the
particles up. This heating is caused by the strong compression
characterizing the second passage of the gas through pericenter. The
compression is caused by the gas orbits converging to the space
occupied by the star at the initial pericenter
\citep[e.g.][]{1994ApJ...422..508K}. This compression can be seen in
Fig.~\ref{fig:press} where we show the pressure at some fixed time
after the first return to pericenter in the PS2 run. The pressure is
shown along a path as a function of $l$, the length of the path. The
path was chosen so as to pass through the gas stream and pericenter.
As can be seen, the pressure rises sharply at negative $l$,
corresponding to the passage through pericenter, and there is a large
pressure gradient in the $z$ direction.  We find that the bounce that
follows this compression is sufficiently strong to impart a
significant fraction of the gas with the escape velocity. In
Fig.~\ref{fig:mb} we show the amount of the star's mass that is
bound, unbound, and accreted (we note again that accreted here means
that it has come closer then $10R_g$ to the BH).  As the star leaves
the vicinity of the BH for the first time, 65\% of its mass is bound.
This stays constant during the gas stream phase since the pressures
are small. Following the second passage through pericenter however,
mass gets unbound, until, at the time we stop our simulation, 50\% of
the mass is unbound, 40\% is bound and the remaining 10\% is accreted.
The relative difference between the two runs is about 10\% in all of
these quantities.

\begin{figure}
  \begin{center}
    \includegraphics[width=12cm]{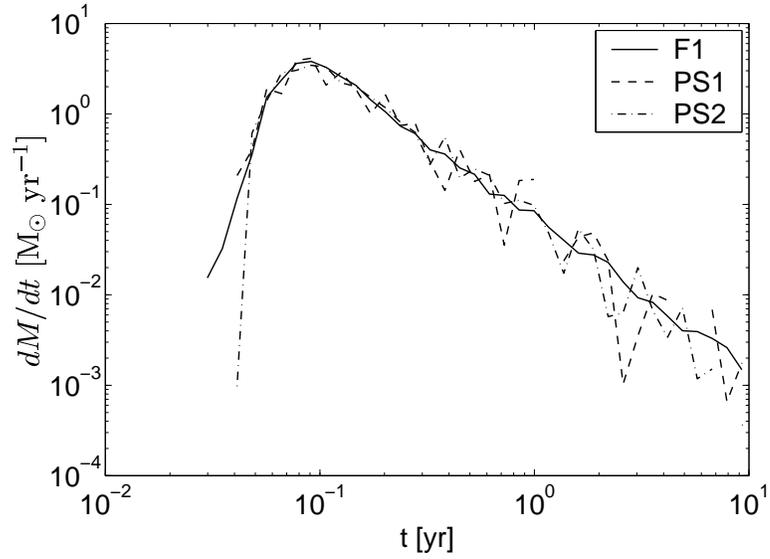}
    \caption{\label{fig:dmdt}
      Differential mass distribution in orbital period for the bound
      mass. The PS runs have less particles at this stage than the F1
      run (870, 2040 and 4295 particles for the PS1, PS2 and F1 runs
      respectively) at this stage so the distribution is less smooth,
      especially in the long period masses. Both runs display a
      similar distribution near the maximum and up to $0.2{\rm ~yr}$,
      when we stop the simulation. }
      \end{center}
\end{figure}

\begin{figure}
  \begin{center}
    \includegraphics[width=10cm]{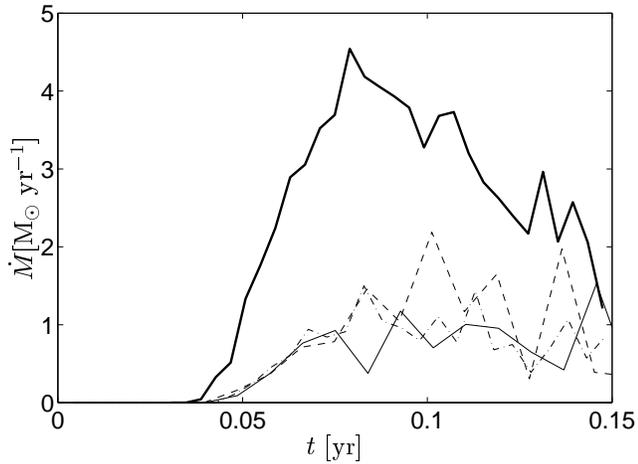}
    \caption{\label{fig:accreted}
      Accretion rates. The solid line is the accretion rate assuming
      that all the returning mass is accreted; the other three lines are 
      the actual accretion rate for the three runs.
      }
  \end{center}
\end{figure}

In Fig.~\ref{fig:dmdt} we show the differential mass distribution in
orbital periods in the stream. This has been used in previous works
\citep[e.g.][]{1988Natur.333..523R,1989ApJ...346L..13E,1993ApJ...410L..83L}
to estimate the mass infall rate onto the BH, under the assumption
that all the mass returning to pericenter is accreted. In
Fig.~\ref{fig:accreted} we show the estimated accretion rate according
to Fig.~\ref{fig:dmdt} together with the \textit{actual} accretion rate we get.
Our calculations show that since some of the mass becomes unbound when
reaching pericenter, the actual accretion rate is a factor of 3
lower than that inferred previously. The total accreted mass up to 60
days (Fig.~\ref{fig:mb}) was overestimated by a factor of 4 when
assuming that all returning mass is accreted. Another important consequence 
of the strong compression near pericenter is that the expected
self-intersection of the gas stream in fact does not occur, since the debris
is given a high velocity perpendicular to the orbital plane.

\begin{figure}
  \begin{center}
    \includegraphics[width=14cm,clip]{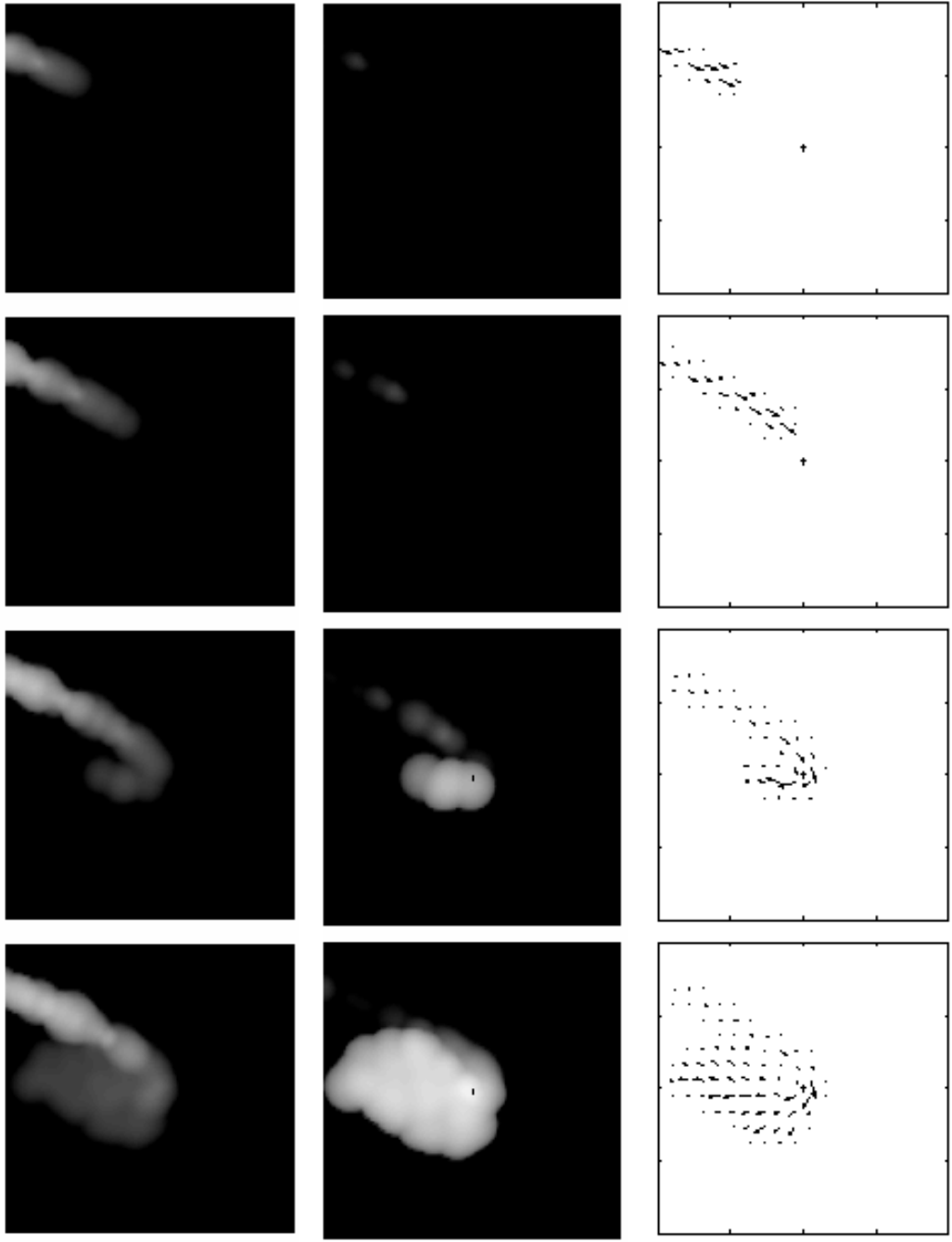}
    \caption{\label{fig:movie1}
      On the left logarithm of the density, in the middle, logarithm
      of the specific internal energy and on the right velocity
      components for various times during the return to pericenter.
      Results are taken from the PS2 run. The BH is at the center, the
      box size is 2000~R$_\odot$. }
  \end{center}
\end{figure}

\begin{figure}
  \begin{center}
    \includegraphics[width=14cm,clip]{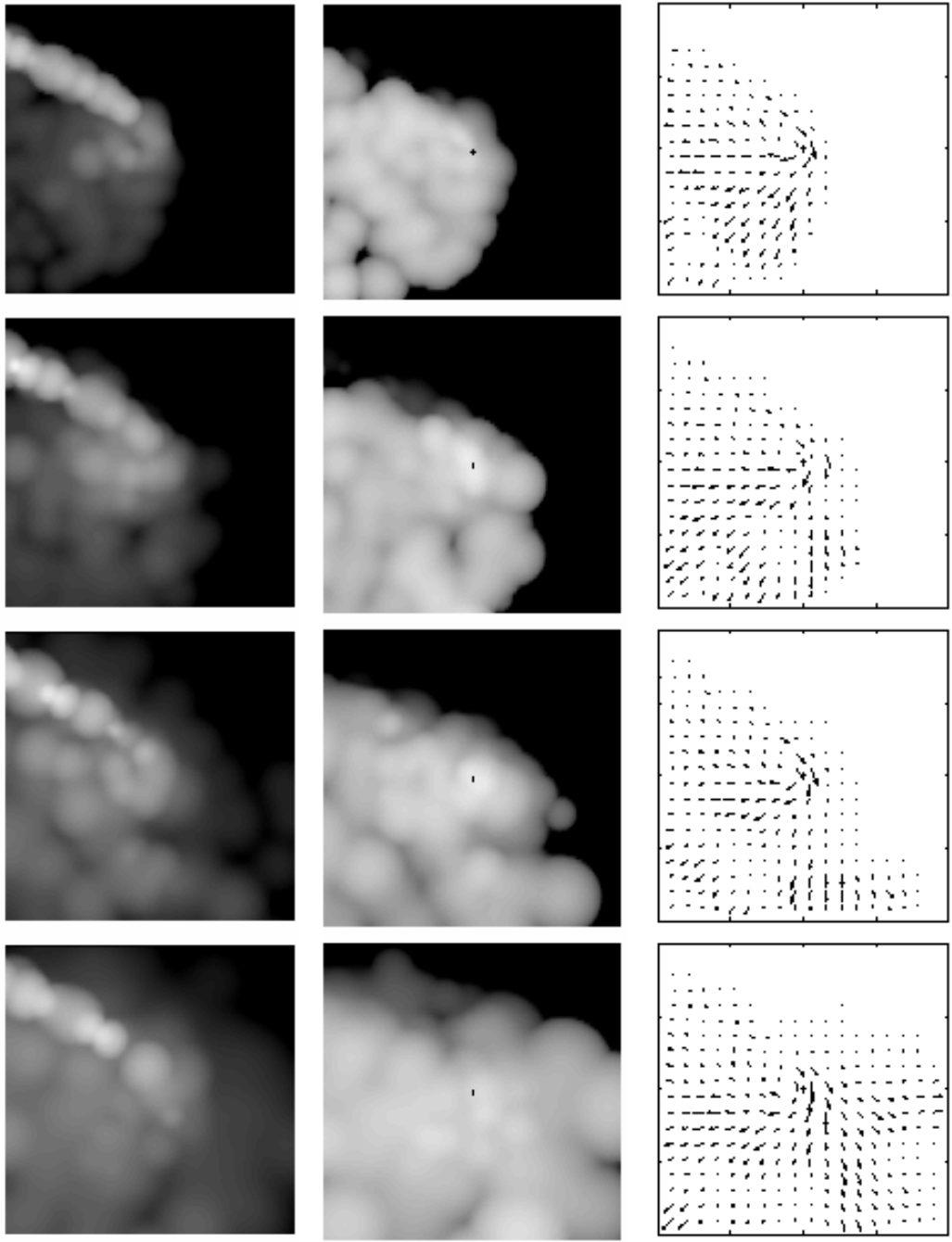}
    \caption{\label{fig:movie2}
      More frames like Fig.~\ref{fig:movie1}.
      }
  \end{center}
\end{figure}

The inner $10^4$~R$_\odot$ around the BH is composed of the infalling
debris and a low density cloud with a density of $\approx 10^{-11}~{\rm
g~cm^{-3}}$ and with a temperature range of higher than $10^6$~K.
In Figs.~\ref{fig:movie1} and~\ref{fig:movie2} we show the density,
specific internal energy and velocity in the inner 2000~R$_\odot$ around
the BH. Even using PS is is clear that we cannot resolve structures
well within this radius (e.g.\ an accretion disk, as proposed by
\citealt{1990ApJ...351...38C}). Nevertheless, there is some evidence of
circularization of the gas flow in the velocity plots.

\section{Discussion}
\label{sec:Conc}
We have performed a 1PN simulation of a solar mass star being disrupted 
by a super massive ($10^6$~M$_\odot$) black hole. The disruption process
itself causes about 1/2 of the star's matter to become unbound. We follow this 
matter up to and beyond the time when it returns to pericenter and starts 
accreting onto the BH. This in turn enables us to determine the amount of 
returning mass that is actually accreted. Contrary to previous, more heuristic 
estimates, we find that only about 25\% of the returning mass actually gets 
accreted. The rest becomes unbound, following being heated by the strong 
compression accompanying the approach to pericenter.

The main consequence of this process is that the maximum accretion
rate is a factor of 3 lower than expected from a simple examination of
the rate of mass return to pericenter. The accretion rate into the
volume resolved by our simulation is also quite constant, at about
1~${\rm M_\odot~yr^{-1}}$, for the last 20~days of the simulation as
opposed to a power law decay which is expected from the rate of
return. When integrated over time, these results show that only about
10\% of the original star's mass actually gets accreted as opposed to
the 65\% expected if all the bound mass would be finally accreted.
This means that the mass involved in the accretion flow, be it in the
form of an accretion disk \citep[e.g.][]{1990ApJ...351...38C}, or
spherical accretion \citep[e.g.][]{Loeb97} is considerably smaller
than previously estimated.  Consequently, the duration of the expected
`flare' can be shorter by a factor 4--5 compared to these early
estimates, which makes the detectability of these events much harder
(the rather high temperature of some of the infalling debris also
makes the bolometric correction high). Indeed, supernova searches in
distant galaxies have failed so far to identify any such event
unambiguously (Filippenko 2000, private communication).

The mass of the debris that becomes unbound because of the pericenter
compression is comparable to the mass of the unbound debris resulting
from the original disruption event. The main difference is that this
new debris component has a different orbital distribution. Most
notably, the velocity has a much larger component in the direction
perpendicular to the orbital plane. This additional component of
unbound high velocity debris could produce interesting consequences
(e.g.\ components like Sgr~A East in the Galactic center) when
colliding with the interstellar medium surrounding the BH
\citep[e.g.][]{1996ApJ...457L..61K}. In particular, the fact that the
mass ejection is more spherically symmetric, makes the event more
similar to a normal supernova. Thus, tidal disruption events may
produce a quite unique signature (in terms of their impact on
surrounding gas), in which two supernova-type events are separated by
a few weeks to a few months, with the first one being very anisotropic
(mass being ejected within a solid angle
$\Omega\apgt16(R_*/R_{\rho})^{1/2}(M_*/M_{\rm bh})^{1/2}$ rad$^2$) and
the second more spherically symmetric.

\acknowledgements
ML acknowledges support from NASA Grant NAG5-6857.



\begin{thebibliography}{37}
\expandafter\ifx\csname natexlab\endcsname\relax\def\natexlab#1{#1}\fi

\bibitem[{{Ayal} {et~al.}(2000){Ayal}, {Piran}, {Oechslin}, {Davies}, \&
  {Rosswog}}]{sphpn}
{Ayal}, S., {Piran}, T., {Oechslin}, R., {Davies}, M.~B., \& {Rosswog}, S.
  2000, \apj, submitted

\bibitem[{{Benz}(1990)}]{benz_rev}
{Benz}, W. 1990, in The Numerical Modeling of Nonlinear Stellar Pulsations:
  Problems and Prospects, ed. J.~R. {Buchelr} (Dordrecht: Kluwer), 269--288

\bibitem[{{Bicknell} \& {Gingold}(1983)}]{1983ApJ...273..749B}
{Bicknell}, G.~V. \& {Gingold}, R.~A. 1983, \apj, 273, 749

\bibitem[{{Blanchet} {et~al.}(1990){Blanchet}, {Damour}, \& {Schaefer}}]{BDS}
{Blanchet}, L., {Damour}, T., \& {Schaefer}, G. 1990, \mnras, 242, 289

\bibitem[{{Cannizzo} {et~al.}(1990){Cannizzo}, {Lee}, \&
  {Goodman}}]{1990ApJ...351...38C}
{Cannizzo}, J.~K., {Lee}, H.~M., \& {Goodman}, J. 1990, \apj, 351, 38

\bibitem[{{Carter} \& {Luminet}(1983)}]{1983A&A...121...97C}
{Carter}, B. \& {Luminet}, J.-P. 1983, \aap, 121, 97

\bibitem[{{Carter} \& {Luminet}(1985)}]{1985MNRAS.212...23C}
{Carter}, B. \& {Luminet}, J.~P. 1985, \mnras, 212, 23

\bibitem[{{Diener} {et~al.}(1997){Diener}, {Frolov}, {Khokhlov}, {Novikov}, \&
  {Pethick}}]{1997ApJ...479..164D}
{Diener}, P., {Frolov}, V.~P., {Khokhlov}, A.~M., {Novikov}, I.~D., \&
  {Pethick}, C.~J. 1997, \apj, 479, 164+

\bibitem[{{Evans} \& {Kochanek}(1989)}]{1989ApJ...346L..13E}
{Evans}, C.~R. \& {Kochanek}, C.~S. 1989, \apjl, 346, L13

\bibitem[{{Frank}(1978)}]{1978MNRAS.184...87F}
{Frank}, J. 1978, \mnras, 184, 87

\bibitem[{{Frolov} {et~al.}(1994){Frolov}, {Khokhlov}, {Novikov}, \&
  {Pethick}}]{1994ApJ...432..680F}
{Frolov}, V.~P., {Khokhlov}, A.~M., {Novikov}, I.~D., \& {Pethick}, C.~J. 1994,
  \apj, 432, 680

\bibitem[{{Fryer} {et~al.}(1999){Fryer}, {Woosley}, {Herant}, \&
  {Davies}}]{1999ApJ...520..650F}
{Fryer}, C.~L., {Woosley}, S.~E., {Herant}, M., \& {Davies}, M.~B. 1999, \apj,
  520, 650

\bibitem[{{Goad} \& {Koratkar}(1998)}]{Goad98}
{Goad}, M. \& {Koratkar}, A. 1998, \apj, 495, 718+

\bibitem[{{Grupe} {et~al.}(1999){Grupe}, {Thomas}, \& {Leighly}}]{Grupe99}
{Grupe}, D., {Thomas}, H.~., \& {Leighly}, K.~M. 1999, \aap, 350, L31

\bibitem[{{Hills}(1988)}]{Hills88}
{Hills}, J.~G. 1988, \nat, 331, 687

\bibitem[{{Kallman} \& {Elitzur}(1988)}]{Kallman88}
{Kallman}, T. \& {Elitzur}, M. 1988, \apj, 328, 523

\bibitem[{{Khokhlov} \& {Melia}(1996)}]{1996ApJ...457L..61K}
{Khokhlov}, A. \& {Melia}, F. 1996, \apjl, 457, L61

\bibitem[{{Khokhlov} {et~al.}(1993{\natexlab{a}}){Khokhlov}, {Novikov}, \&
  {Pethick}}]{1993ApJ...418..181K}
{Khokhlov}, A., {Novikov}, I.~D., \& {Pethick}, C.~J. 1993{\natexlab{a}}, \apj,
  418, 181+

\bibitem[{{Khokhlov} {et~al.}(1993{\natexlab{b}}){Khokhlov}, {Novikov}, \&
  {Pethick}}]{1993ApJ...418..163K}
---. 1993{\natexlab{b}}, \apj, 418, 163+

\bibitem[{{Kochanek}(1992)}]{1992ApJ...385..604K}
{Kochanek}, C.~S. 1992, \apj, 385, 604

\bibitem[{{Kochanek}(1994)}]{1994ApJ...422..508K}
---. 1994, \apj, 422, 508

\bibitem[{{Komossa} \& {Bade}(1999)}]{Komossa99a}
{Komossa}, S. \& {Bade}, N. 1999, \aap, 343, 775

\bibitem[{{Komossa} \& {Greiner}(1999)}]{Komossa99b}
{Komossa}, S. \& {Greiner}, J. 1999, \aap, 349, L45

\bibitem[{{Lacy} {et~al.}(1982){Lacy}, {Townes}, \&
  {Hollenbach}}]{1982ApJ...262..120L}
{Lacy}, J.~H., {Townes}, C.~H., \& {Hollenbach}, D.~J. 1982, \apj, 262, 120

\bibitem[{{Laguna} {et~al.}(1993){Laguna}, {Miller}, {Zurek}, \&
  {Davies}}]{1993ApJ...410L..83L}
{Laguna}, P., {Miller}, W.~A., {Zurek}, W.~H., \& {Davies}, M.~B. 1993, \apjl,
  410, L83

\bibitem[{{Loeb} \& {Ulmer}(1997)}]{Loeb97}
{Loeb}, A. \& {Ulmer}, A. 1997, \apj, 489, 573+

\bibitem[{{Marck} {et~al.}(1996){Marck}, {Lioure}, \& {Bonazzola}}]{Marck96}
{Marck}, J.~A., {Lioure}, A., \& {Bonazzola}, S. 1996, \aap, 306, 666+

\bibitem[{{Monaghan}(1992)}]{mon_rev}
{Monaghan}, J.~J. 1992, \araa, 30, 543

\bibitem[{{Nolthenius} \& {Katz}(1982)}]{1982ApJ...263..377N}
{Nolthenius}, R.~A. \& {Katz}, J.~I. 1982, \apj, 263, 377

\bibitem[{{Peterson} \& {Ferland}(1986)}]{Peterson86}
{Peterson}, B.~M. \& {Ferland}, G.~J. 1986, \nat, 324, 345

\bibitem[{{Press} \& {Teukolsky}(1977)}]{1977ApJ...213..183P}
{Press}, W.~H. \& {Teukolsky}, S.~A. 1977, \apj, 213, 183

\bibitem[{{Rees}(1988)}]{1988Natur.333..523R}
{Rees}, M.~J. 1988, \nat, 333, 523

\bibitem[{{Rees}(1990)}]{1990Sci...247..817R}
---. 1990, Science, 247, 817

\bibitem[{{Renzini} {et~al.}(1995){Renzini}, {Greggio}, {di Serego-Alighieri},
  {Cappellari}, {Burstein}, \& {Bertola}}]{Renzini95}
{Renzini}, A., {Greggio}, L., {di Serego-Alighieri}, S., {Cappellari}, M.,
  {Burstein}, D., \& {Bertola}, F. 1995, \nat, 378, 39+

\bibitem[{{Storchi-Bergmann} {et~al.}(1995){Storchi-Bergmann}, {Eracleous},
  {Livio}, {Wilson}, {Filippenko}, \& {Halpern}}]{1995ApJ...443..617S}
{Storchi-Bergmann}, T., {Eracleous}, M., {Livio}, M., {Wilson}, A.~S.,
  {Filippenko}, A.~V., \& {Halpern}, J.~P. 1995, \apj, 443, 617

\bibitem[{{Syer} \& {Clarke}(1992)}]{1992MNRAS.255...92S}
{Syer}, D. \& {Clarke}, C.~J. 1992, \mnras, 255, 92

\bibitem[{{Syer} \& {Clarke}(1993)}]{1993MNRAS.260..463S}
---. 1993, \mnras, 260, 463+

\bibitem[{{Terlevich} \& {Melnick}(1988)}]{Terlevich88}
{Terlevich}, R. \& {Melnick}, J. 1988, \nat, 333, 239+

\end{thebibliography}
\end{document}